\documentstyle[11pt,newpasp,twoside,epsf]{article}
\def \la {\mathrel{\vcenter
     {\offinterlineskip \hbox{$<$}\hbox{$\sim$}}}}
\def \ga {\mathrel{\vcenter
     {\offinterlineskip \hbox{$>$}\hbox{$\sim$}}}}

\def\ave#1{\langle #1 \rangle}

\def\edcomment#1{\iffalse\marginpar{\raggedright\sl#1\/}\else\relax\fi}
\reversemarginpar

\def\ave#1{\langle #1 \rangle}
\begin{document}
\title{Detectable Signals from Mergers of Compact Stars}
 \author{Hans-Thomas Janka}
\affil{Max-Planck-Institut f\"ur Astrophysik, Karl-Schwarzschild-Str.~1,
D-85741 Garching, Germany}
\author{Maximilian Ruffert}
\affil{Department of Mathematics \& Statistics, University of Edinburgh,
Edinburgh, EH9~3JZ, Scotland, U.K.}

\begin{abstract}
Merging neutron stars and neutron star--black holes binaries are
powerful sources of gravitational waves. They have also been suggested
as possible sources of cosmic gamma-ray bursts and are discussed as
sites for the formation of r-process elements. Whereas the first aspect
is undoubtable, the latter two are very uncertain. The current status of 
our knowledge and of the 
numerical modeling is briefly reviewed and the results of simulations are 
critically discussed concerning their significance and implications for
potentially observable signals.
\end{abstract}

\section{Introduction}

Mergers of binary neutron stars (NS+NS) and neutron stars with companion
black holes (NS+BH) are known to occur, because the emission of gravitational
waves leads to a shrinking orbital separation and does not allow the
systems to live forever. The Hulse-Taylor binary pulsar PSR~B1913+16 
(Hulse \& Taylor 1975) is
the most famous example among a handful of known systems (Thorsett \&
Chakrabarty 1999).
Eventually, after 10$^8$--$10^{10}$ years of evolution, these double stars
will approach the final catastrophic plunge and will become powerful sources
of gravitational radiation. This makes them one of the most promising targets
for the upcoming huge interferometer experiments, being currently under
construction in Europe (GEO600, VIRGO), Japan (TAMA) and the United States
(LIGO). The expected rate of merging events is of the order of $10^{-5}$
per year per galaxy, estimated, with significant uncertainties, 
from population synthesis models and empirical data (see, e.g., 
Bulik, Belczynski, \& Zbijewski 1999; Fryer, Woosley, \& Hartmann 1999; 
Kalogera, \& Lorimer 2000; and references therein). Templates are urgently
needed to be able to extract the expected signals from background noise
and to interpret the meaning of possible measurements.

Merging compact objects were suggested as possible sources of gamma-ray
bursts (Blinnikov et al. 1984; Eichler et al. 1989; Paczy\'nski 1991;
Narayan, Piran, \& Shemi 1991), but
so far observations cannot provide convincing arguments for this 
hypothesis. In fact, the exciting detection of afterglows at other wavelengths
of the electromagnetic spectrum by the Dutch-Italian {\em BeppoSAX} satellite
and the discovery of lines in the afterglow spectra support the association
of gamma-ray bursts (GRBs) with explosions of massive stars in star-forming 
regions of galaxies at high redshifts (e.g., Klose 2000, Lamb 2000). However, 
so far afterglows were discoverd only for GRBs with durations longer than a few
seconds. Similar observations for the class of short, hard bursts ($\Delta
t_{\mathrm{GRB}} \la 2$~seconds; Mao, Narayan, \& Piran 1994) 
are still missing, and we have neither a 
proof of their origin from cosmological distances, nor any hint on the 
astrophysical object they are produced by. NS+NS and NS+BH mergers are 
therefore still viable candidates for at least the subclass of short-duration
bursts, in particular because such collisions of compact objects must occur at 
interesting rates, can release huge amounts of energy, and the energy is set
free
in a volume small enough to account for intensity fluctuations on a millisecond 
timescale. The ultimate proof of such an association would certainly be the 
coincidence of a GRB with a characteristic gravitational-wave signal.

Lattimer \& Schramm (1974, 1976), 
Lattimer et al. (1977), and Meyer (1989) first considered merging neutron stars
as a possible source of r-process nuclei. Their idea was that some matter
might get unbound during the dynamical interaction. Since in neutron star
matter neutrons are present with high densities and heavy nuclei can exist
in surface-near layers, the environment might be suitable for the formation
of neutron-rich, very massive nuclei. Detailed models are needed to answer
the question how much matter can possibly be ejected during the merging,
and what the nuclear composition of this material is.

Doing hydrodynamic simulations of the merging of compact stars is a 
challenging problem. Besides three-dimensional hydrodynamics, preferably
including general relativity, the microphysics within the neutron stars
is very complex. A nuclear equation of state (EoS) has to be used, neutrino 
physics may be important when the neutron stars heat up, magnetic 
fields might cause interesting effects, etc. 
Also high numerical resolution is required
to account for the steep density gradient at the neutron star surface, and
a large computational volume is needed if the ejection of matter is to
be followed. 

Different approximations and simplifications are possible when
different aspects are in the main focus of interest. For example, the surface
layers are unimportant and a simple ideal gas EoS [$P = (\gamma-1)\varepsilon$]
and polytropic structure of the neutron stars can be
used for parametric studies, when the gravitational-wave emission is to
be calculated. This is an absolutely unacceptable simplification, however,
when mass ejection and nucleosynthesis shall be investigated.
For the latter problem as well as for the GRB topic, on the other hand,
it is not clear whether general relativistic physics is essential. Nevertheless,
of course, general relativity may be important to get quantitatively meaningful 
results. In particular, Newtonian models cannot answer the question whether 
and when a black hole is going to form after two neutron stars have merged.

Remarkable progress has been achieved during the past years.
General relativistic simulations are in reach now, at least with simple
input physics and for certain phases of the evolution of the binary
systems, especially for the inspiraling phase (Duez, Baumgarte, \& Shapiro 2000)
and for the first stage of the final dynamical plunge 
(Shibata \& Ury\=u 1999; Ury\=u, Shibata, \& Eriguchi 2000), although
the simulations become problematic once an apparent horizon begins to form.
On the other hand, first attempts have been made to add detailed microphysics,
i.e., a physical nuclear EoS and neutrino processes, into Newtonian 
or to some degree post-Newtonian models (Ruffert, Janka, \& Sch\"afer 1996;
(Ruffert et al. 1997; Ruffert \& Janka 1998, 1999; Rosswog et al. 1999, 2000). 
The hydrodynamics results were used in post-processing steps to draw
conclusions on implications for GRB scenarios (Janka \& Ruffert 1996,
Ruffert \& Janka 1999, Janka et al.\ 1999) and to evaluate the
density-temperature trajectories of ejected matter for nucleosynthesis 
processes (Freiburghaus, Rosswog, \& Thielemann 1999). 
The results, however, are not finally 
conclusive, because important effects have so far not been fully included and
the simulations have to be improved concerning the resolution of the
neutron star surfaces to reduce associated numerical artifacts.

\begin{figure}
\plotone{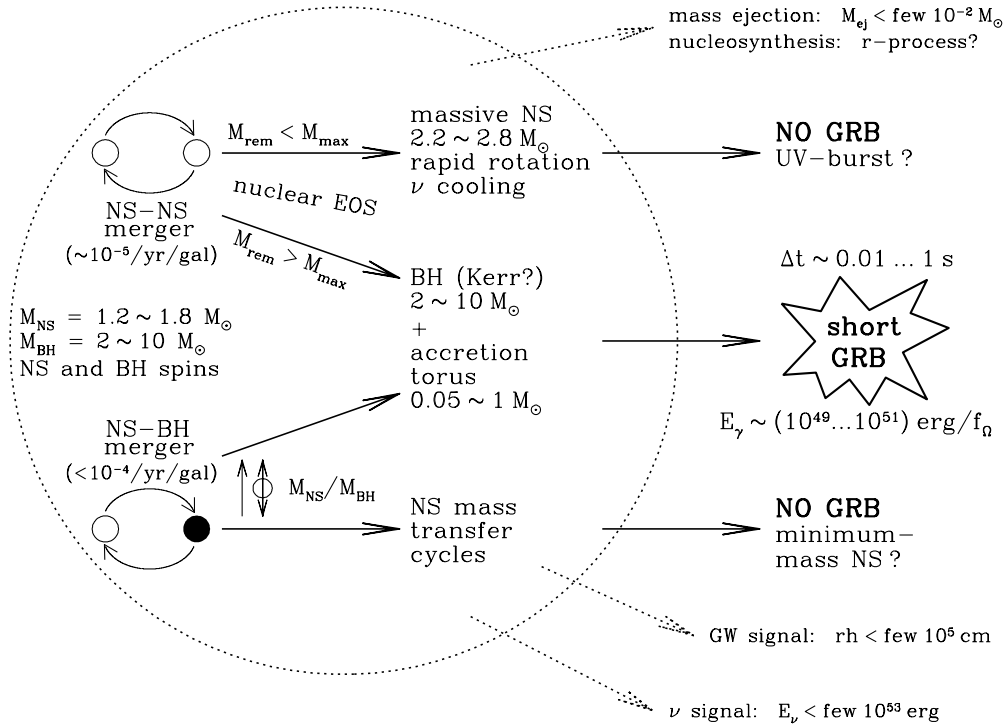}
\caption{\small
Possible evolution and observable signatures of NS+NS and BH+NS mergers. 
The evolution depends on
the properties of the merging objects and the unknown conditions in the
supranuclear interior of the neutron stars.}
\end{figure}

\section{Importance of the Nuclear Equation of State}

Independent of numerical aspects and the unsatisfactory status of including
general relativistic effects, one of the handicaps of merger models
of compact stars are the unknown properties of the nuclear EoS. The
compressibility of the nuclear EoS determines the neutron star structure and
mass-radius relation. Therefore on the one hand, the evolution right before 
the merging, Roche-lobe overflow and the details of the final orbital 
instability and plunge of NS+NS binaries are affected by the neutron star EoS.
On the other hand the possibility of episodic mass transfer and of cycles of 
orbital decay and widening in BH+NS systems depend on the EoS properties
(Lattimer \& Prakash 2000a; Lee \& Klu\'zniak 1999a,b;
Lee 2001). The nuclear EoS thus determines the dynamics and
gravitational-wave emission during these phases. But also the post-merging 
evolution is very sensitive to the EoS properties.
The NS+NS merger remnant can be a rapidly and
differentially rotating, protoneutron-star like object. If the
EoS is stiff, this object might not collapse to a black hole on a dynamical
timescale, but its mass may be below the maximum stable mass of such a 
configuration. In this case the merger remnant might approach the final
instability only on a much longer secular timescale, driven by neutrino 
emission, viscous and magnetic braking (Lipunova \& Lipunov 1998)
and angular momentum loss due to gravitational wave emission or mass
shedding (Baumgarte \& Shapiro 1998). 

This could have very important implications for the gravitational-wave
signal, because the emission might not be shut off immediately after the 
collision of the binary components, but --- in particular in case of possibly
developing triaxiality of the post-merger object --- might continue for a 
considerable period of time. On the other hand, a hot, stable, massive 
protoneutron star-like object would cool by neutrino emission. Since neutrino
energy deposition in the surface layers of the remnant causes a baryonic outflow
of matter, similar to what happens in case of a protoneutron star at the center
of a supernova, the merger site would be ``polluted'' by baryons and 
gamma-ray burst models might have a problem to explain the ejection of a baryon-poor 
pair-plasma fireball with highly relativistic velocities (Lorentz factors of 
several 100), which is needed to produce the visible gamma-ray emission.
Of course, ways have been suggested to reduce or circumvent 
this problem, for example by considering a non-standard composition (strange quark
matter) and structure (no baryonic crust, extreme compactness) of the merger 
remnant (e.g., Bombaci \& Datta 2000; Mitra 2000; and references therein), or
by invoking magnetic fields at the engine of the gamma-ray burst (e.g., 
Klu\'zniak \& Ruderman 1998; Usov 1994).

The currently discussed EoSs for the supranuclear regime
do not set strict limits for the maximum mass of neutron stars.
Whereas the nuclear EoS around saturation density is expected to be
rather stiff with a neutron star mass limit around 2~M$_{\odot}$ or 
larger, there is very little known about the EoS properties above twice
nuclear matter density. A possible phase transition to a hyperonic state,
kaon or pion condensates or a quark phase might soften the EoS
and reduce the mass limit to significantly lower values 
(Lattimer \& Prakash 2000b; Heiselberg \& Pandharipande 2000; Balberg \& Shapiro 2000;
Weber 1999). Rigid rotation, on the other hand, can increase this limit by at
most $\sim 20$\%, but differential rotation, which is likely to be present
in the post-merger configuration, can stabilize compact stars with much larger
masses (Baumgarte, Shapiro, \& Shibata 2000). The degree of differential rotation 
in the remnant of a neutron star merger can only be determined by hydrodynamic
simulations. On the other hand, the post-merging object
is very massive (its baryonic mass is roughly twice the 
mass of a single neutron star), and due to the violence of the dynamical interaction
of the merging stars the equilibrium density is overshot, thus favoring 
the gravitational collapse on a dynamical timescale even when an object with
the combined mass of both neutron stars could find a stable rotational
equilibrium.

Therefore firm theoretical predictions about the post-merging evolution of 
NS+NS systems are not possible at the moment, not even in case of a fully 
general relativistic treatment. Figure~1 summarizes possible 
evolutionary scenarios in dependence of the EoS properties and the masses
and spins of the merging compact stars. When NS+NS and BH+NS merging leads to
a (rapidly rotating) black hole surrounded by an accretion torus,
energy can be released on the accretion timescale of the gas, which is 
100 to 1000 times longer than the dynamical timescale of the system. In this
case thermal energy of the hot, dense torus can be radiated through neutrinos, 
or rotational energy of the disk or black hole might be extracted by magnetic
fields (Blandford \& Znajek 1971; Li \& Paczy\'nski 2000).
The efficiency of energy conversion to
neutrinos or Poynting flux can be very high: A fair fraction of the rest mass
energy of the accreted gas (up to several 10 per cent)
or the rotational energy of the black hole can be released 
that way. Numerical values depend on the mass and the spin angular momentum 
of the black hole, on the strength of the magnetic fields and on the physical 
conditions in the accretion torus. Since the outflows occur preferentially
along the rotational axis of the system, one hopes that the baryon loading 
problem can be avoided by the lower baryon densities above the 
poles of the black hole. Detailed models are necessary to make
quantitative statements. Hydrodynamical simulations have been performed
for the neutrino-powered scenario, although problems like the baryonic
entrainment of the outflow and the lifetime of the accretion torus
are still unanswered and require further refinement of the models. 
In case of the magnetohydrodynamic mechanism time-dependent simulations 
are extremely challenging and are currently not at the horizon.

Once bipolar outflow has formed, the question remains to be answered, 
what fraction of the kinetic or magnetic energy can finally be converted
into gamma rays. Pessimistic estimates quote a few per cent or less, 
optimistic values claim several 10\% (Kobayashi \& Sari 2000).

Provided the remnant of a neutron star merger does not collapse to a black
hole on a dynamical timescale, the baryon pollution problem is serious
and probably hampers or destroys the chance for a gamma-ray burst. A massive,
hot neutron star, which cools by neutrino emission, will poison its surroundings
with a considerable amount of baryonic matter. Neutrino-driven winds might 
carry off $\sim 10^{-2}$~M$_{\odot}$ or more in all directions. This could 
have observable implications of such events, possibly generating a UV outburst
of radiation when the expanding cloud dilutes to reach transparency (Li \&
Paczy\'nski 1998). Concerning baryon pollution BH+NS mergers certainly
offer an advantage because of the permanent presence of the black hole. 
The baryon pollution problem, however, might also occur there
if the system runs into a state of long lasting, episodic mass transfer and 
the donating neutron star is not quickly disrupted into a disk. Circling around
the black hole on a close orbit will lead to tidal heating of the neutron 
star and corresponding neutrino emission. In case of mass stripping of the 
neutron star down to its minimum mass, an explosion may occur, again possibly
associated with the emission of radiation which is powered by radioactive 
decays in the expanding neutron star matter (Sumiyoshi et al.\ 1998).

\begin{figure}
\vspace{17truecm}
\caption{\small
Phases of mass ejection from NS+NS and BH+NS mergers.}
\end{figure}

\section{Nucleosynthesis in the Ejecta}

Apart from a possible explosion of a neutron star at its minimum stable mass, 
two distinct 
processes can contribute to mass ejection from merger systems (Fig.~2).
On the one hand, there is a phase of dynamical ejection of cold, low-entropy
surface material of the merging neutron stars, which starts
shortly after the two stars have fallen into each other and the neutron
stars develop long tidal arms by centrifugal forces. The amount of mass loss
depends very sensitively on the total angular momentum of the system and
is largest when the neutron star spin(s) is (are) aligned with the orbital angular
momentum vector. Decompressed neutron star surface matter expands and releases 
energy through radioactive decays or nucleon recombination to nuclei, which might 
help the flow to become 
gravitationally unbound. On the other hand, general relativity, if it leads
to a quick collapse of the merger remnant to a black hole, might suppress
the possibility of ejecting material from the tips of the tidal tails.

A second phase of mass loss is connected with the neutrino-driven wind. The matter
will be neutrino-heated and hot, i.e., the entropies are very high. This 
plasma outflow is unavoidable when the neutron stars heat up and emit neutrinos,
or when a hot, radiating accretion torus is swallowed by a black hole after the 
merging.

Nucleosynthesis calculations for the dynamically ejected matter of 
NS+NS mergers have been attempted
(Freiburghaus et al.\ 1999), but the problem is not easy to attack.
Not only the disregard of general relativity implies uncertainties, also the
numerical resolution of the surface layers of the neutron stars has to be 
good enough to calculate the thermodynamical conditions precisely. The   
current models yield sufficiently large temperatures ($T \ga 5\times 10^9$~K)
that the matter reaches nuclear statistical equilibrium (NSE) and the r-processing 
proceeds very similar to the classical r-process in the shock-heated, expanding
matter of supernovae. With a suitable superposition of conditions (degree of
neutronization for certain amounts of ejected matter) the integral abundance curve
can be forced to fit the solar r-process abundance distribution. But what is 
the ``actual'' (model determined) neutronization of the ejecta?
The effects of radioactive decays and neutrino production on the neutron density
have to be taken into account to answer this question. 
And, also crucial, what are the ``true'' temperatures in the ejected gas?
It is not certain that the gas in the very extended spiral arms ever reaches
NSE temperature. If it stays cooler, the neutron-rich isotopes which are
present in the outer layers of the neutron star, would be swept into the 
interstellar medium and after decaying, would reflect an abundance distribution 
different from the solar r-process pattern. In hydrodynamical models
one has to fight against numerical viscosity when the temperature is to be
calculated very accurately.

There is plenty of space for future work. Due to the lack of detailed
models, nucleosynthesis has not been studied for the anisotropic neutrino-driven
outflow of accreting black holes as merger remnants. In fact, the
environment might provide suitable conditions of entropy and neutronization to
obtain high-entropy r-processing. Such conditions can be verified in 
neutrino-driven winds of protoneutron stars in supernovae only in extreme
corners of the parameter space for neutron star masses and radii 
(Qian \& Woosley 1996; Otsuki et al.\ 2000; Sumiyoshi et al.\ 2000).

{\small
\begin{table}[t]
\tabcolsep 4pt
\begin{center}
\caption{NS+NS and BH+NS merger simulations}
\begin{tabular}{lcccccccc}
\\
\hline\hline\\[-3mm]
Model & Type & Masses & Spin & $t_{\rm sim}$ & $t_{\rm ns}$ & $d_{\rm ns}$ 
& $M_{\rm ns}^{\rm min}$ & $\Delta M_{\rm ej}$\\[1mm]
   &   & $M_{\odot}$ &  & ms & ms & km & $M_{\odot}$ &
$M_{\odot}/100$\\[1mm]
\hline\\[-3mm]
   S64   & NS+NS & 1.2+1.2  & solid & 10 & 2.8 & 15 & ... & 2.0   \\
   D64   & NS+NS & 1.2+1.8  & solid & 13 & 7.3 & 15 & ... & 3.8   \\
   C64   & NS+NS & 1.6+1.6  & anti  & 10 & 3.7 & 15 & ... & 0.0085\\
   A64   & NS+NS & 1.6+1.6  & none  & 10 & 1.7 & 15 & ... & 0.23  \\
   B64   & NS+NS & 1.6+1.6  & solid & 10 & 1.6 & 15 & ... & 2.4   \\
\hline
   TN10  & BH/AD & 2.9+0.26 & solid & 15 & ... & ... & ... & ...  \\
\hline
   C2.5  & BH+NS & 2.5+1.6  & anti  & 10 & 2.6  & 11 & 0.78 & 0.01\\
   A2.5  & BH+NS & 2.5+1.6  & none  & 10 & 4.3  & 18 & 0.78 & 0.03\\
   B2.5  & BH+NS & 2.5+1.6  & solid & 10 & 6.0  & 23 & 0.78 & 0.2 \\
   C5    & BH+NS & 5.0+1.6  & anti  & 15 & 9.1  & 76 & 0.40 & 2.5 \\
   A5    & BH+NS & 5.0+1.6  & none  & 20 & 16.3 & 65 & 0.52 & 2.5 \\
   B5    & BH+NS & 5.0+1.6  & solid & 15 & 10.8 & 79 & 0.50 & 5.6 \\
   C10   & BH+NS & 10.0+1.6 & anti  & 10 & 8.0  & 96 & 0.65 & 2.2 \\
   A10   & BH+NS & 10.0+1.6 & none  & 10 & 9.3  & 95 & 0.60 & 3.2 \\
   B10   & BH+NS & 10.0+1.6 & solid & 10 & 5.1  & 97 & 0.65 & 10.0\\[1mm]
\hline
\end{tabular}
\end{center}
\end{table}
}

\section{Computational Methods and Computed Models}

Here we will only summarise the computational procedures that we used
for doing NS+NS and BH+NS merger simulations.
Details of the hydrodynamic method as well as the neutrino relevant
algorithms can be found in Ruffert et al.\ (1996, 1997).
The nested grid was described by Ruffert \& Janka~(1998) and
Ruffert~(1992). 
The implementation of the black hole was explained in Ruffert \&
Janka~(1999), Eberl~(1998) and Janka et al.~(1999). 

\subsection{Methods}

The hydrodynamical simulations were done with a code based on the
Piecewise Parabolic Method (PPM) developed by Colella \& Woodward~(1984).
The code is basically Newtonian, but contains the terms 
necessary to describe gravitational-wave emission and the corresponding
back-reaction on the hydrodynamical flow (Blanchet et al~1990).
The modifications that follow from the gravitational
potential are implemented as source terms in the PPM algorithm,
which are obtained from fast Fourier transforms.
The necessary spatial derivatives are evaluated as standard centered
differences on the grid. 

In order to describe the thermodynamics of the neutron star matter, 
we use the EoS of Lattimer \& Swesty (1991)
for a compressibility modulus of bulk nuclear matter of
$K = 180\,$MeV in tabular form. Use of a physical EoS instead of a 
simple ideal gas law implies that the adiabatic index is a function
of space (i.e., of density, temperature and composition in the star) 
as well as of time (i.e., the value of the adiabatic index of the 
bulk of the matter changes when the neutron star strips some of its
mass). This is an important
difference compared to the polytropic neutron stars considered by
Lee (2001). As mentioned above, the effective adiabatic index
influences the dynamics of NS+NS and BH+NS mergers. An ideal gas 
description defines an idealized, ``pure'' condition, the ``realistic'' 
case is more complex.

Energy loss and changes of the electron
abundance due to the emission of neutrinos and antineutrinos are 
taken into account by an elaborate ``neutrino leakage scheme''.
The energy source terms contain the production of all types of 
neutrino pairs by thermal processes and additionally of electron 
neutrinos and antineutrinos by lepton captures onto baryons. The latter
reactions act as sources or sinks of lepton number, too, and are
included as source terms in a continuity equation for the electron
lepton number. Matter is rendered optically thick to neutrinos due to
the main opacity producing reactions, which are neutrino-nucleon
scattering and absorption of electron-type neutrinos onto nucleons.

The presented simulations were done on multiply nested and refined grids.
With an only modest increase in CPU time, the nested grids allow one to
simulate a substantially larger computational volume while at the same
time they permit a higher local spatial resolution of the merging
stars. The former is important to follow the fate of matter that is
flung out to distances far away from the merger site either to
become unbound or to eventually fall back. The latter is necessary to
adequately resolve the strong shock fronts and steep discontinuities 
of the plasma flow that develop during the collision.
The procedures used here are based on the algorithms that can be found
in Berger \& Colella~(1989), Berger~(1987) and Berger \& Oliger~(1984).
Each grid had 64$^3$ zones, the size of the smallest zone was 
$\Delta x = \Delta y = \Delta z =$ 0.64 or 0.78$\,$km in case of binary
neutron stars and 1.25 or 1.5$\,$km for BH+NS
mergers. The zone sizes of the next coarser grid levels were doubled
to cover a volume of 328 or 400$\,$km side length 
for NS+NS and 640 or 768$\,$km for BH+NS simulations.

In a post-processing step, performed after the hydrodynamical evolution
had been calculated, we evaluated our models for neutrino-antineutrino
($\nu\bar{\nu}$) annihilation in the surroundings of the merged stars
in order to construct maps showing the local energy deposition rates
per unit volume. Spatial integration finally yields the total rate of
energy deposition outside the neutrino emitting high-density regions.

\subsection{Models and Initial Conditions}

NS+NS merger simulations were started with two identical Newtonian
neutron stars, or with neutron stars of different mass. In case of 
BH+NS coalescence we only considered neutron stars with a baryonic
mass of about 1.63~$M_\odot$ so far, and varied the black hole mass.
The distributions of density $\rho$ and electron fraction
$Y_e\equiv n_e/n_b$ (with $n_e$ being the number density of electrons
minus that of positrons, and $n_b$ the baryon number density) were
taken from a one-dimensional model of a cold, deleptonized (neutrino-less)
neutron star in hydrostatic equilibrium and were the same as in
Ruffert et al.~(1996). The initial central temperature was set to
a value of typically a few MeV, the temperature profile decreasing
towards the surface such that the thermal energy was much smaller
than the degeneracy energy of the matter.

For numerical reasons the
surroundings of the neutron stars cannot be treated as completely
evacuated. The density of the ambient medium was set to less than
$10^8$~g/cm$^3$, more than six orders of magnitude smaller than
the central densities of the stars. The total mass on the whole
grid, associated with this finite background density is less than
$10^{-3}\,M_{\odot}$.

We prescribed the orbital velocities of the coalescing neutron stars
according to the motions of point masses, as computed
from the quadrupole formula.
The tangential components of the velocities of the neutron star
centers correspond to Kepler velocities on circular orbits,
while the radial velocity components reflect the emission of
gravitational waves leading to the inspiral of the orbiting bodies.

A spin of the neutron stars around their respective
centers was added to the orbital motion and was varied from model 
to model: The ``A'' models do not have any additional spins added
on top of their orbital velocity.
In this case all parts of the neutron stars start out with the same
absolute value of the velocity, also called `irrotational' motion.
The angular velocity (both magnitude and direction) of the spin in
``B'' models is equal to the angular velocity of the orbit.
This results in a `solid body' type (tidally locked) motion, also 
called `corotating'. ``C'' models represent 
counter-rotating cases where spins and orbital angular
momentum vectors have opposite directions.

We do not relax the neutron stars to equilibrium states before we 
start the simulations. Instead, we set the initial distances to sufficiently
large values that most of the induced oscillations have been damped away
by numerical viscosity before the dynamical phase of the merging is reached.
Test calculations with even larger distances do
not exhibit significant differences of the evolution. We are therefore 
pretty confident that the final orbital instability due to the finite 
size of the objects was not affected by the non-equilibrated initial state.

Table~1 gives a list of computed NS+NS and BH+NS merger models, for
which the physical parameters of the systems were varied ($t_{\mathrm{sim}}$
is the time interval covered by the simulation, $t_{\mathrm{ns}}$ the time
when the two density maxima of the neutron stars are
a stellar radius, i.e., $d_{\rm ns} = 15\,$km, apart, and
$\Delta M_{\mathrm{ej}}$ the amount of dynamically ejected matter).
Besides the baryonic mass of the neutron star(s) and the mass of the
black hole, the spins of the neutron stars were varied. ``Solid''
means synchronously rotating stars, ``none'' irrotational
cases, and ``anti'' counter-rotation, i.e., spin angular momenta
opposite to the direction of the orbital angular momentum vector.
In addition to the listed models, we performed simulations where we 
changed numerical aspects, for example the resolution, initial distances
and initial temperatures, or used an entropy equation to follow the 
temperature evolution. In particular, the stability of the numerical 
handling, the smallness of the effects due to numerical viscosity and 
finite resolution, and the quality of energy and angular momentum conservation
were confirmed by BH+NS runs, where we followed the decay of the orbit
for a large number ($\sim 10$) of full revolutions.

The cool neutron stars with baryonic masses of 1.2, 1.63 or 1.8$\,$M$_{\odot}$
have a radius around 15$\,$km (Ruffert et al.\ 1996),
the radius increasing weakly with the stellar mass.
The runs in Table~1 were started with a center-to-center distance of
42--46$\,$km for NS+NS and with 47$\,$km in case of BH+NS for
$M_{\rm BH} = 2.5\,$M$_{\odot}$, 57$\,$km for
$M_{\rm BH} = 5\,$M$_{\odot}$ and
72$\,$km for $M_{\rm BH} = 10\,$M$_{\odot}$.
The simulations were stopped at a time
$t_{\rm sim}$ between 10$\,$ms and $20\,$ms. The black hole was
treated as a  point mass at the center of a sphere with radius
$R_{\rm s} = 2GM_{\rm BH}/c^2$
which gas could enter unhindered. Its mass and momentum were
updated along with the accretion of matter. 

Model TN10, which is added
for comparison, is a continuation of the NS+NS merger Model~B64, where
at time $t = 10\,$ms the formation of a black hole with a mass
of 2.5$\,$M$_{\odot}$ was assumed, and the accretion of the remaining
gas on the grid ($\sim 0.7\,$M$_{\odot}$) was followed for another 5$\,$ms
until a quasi-steady state was reached. Different times for the black hole 
collapse were tested ($\sim 1$, 2, 3 and 9 ms after the neutron stars had
merged to one body), and gave qualitatively similar results
(Ruffert \& Janka 1999).

\begin{figure}
\vspace{17truecm}
\caption{\small 
Density contours in the orbital plane for the NS+NS merger
Models A64 (left column) and B64 (right column). The density (in g/cm$^3$)
is shown logarithmically, the contours being spaced with intervals of 
0.5~dex. The arrows indicate the velocity field. The time is given in 
the upper right corner of the plots.} 
\end{figure}

\begin{figure}
\vspace{7truecm}
\caption{\small
Contour plots of the Newtonian Model~TN10
in the equatorial plane (left) and perpendicular to it
(right) near the end of the simulation.
The density is displayed together with the velocity field. The density
is measured in g~cm$^{-3}$ and its contours are spaced logarithmically
with intervals of 0.5~dex. The time elapsed from the beginning of the
simulation is about 15~ms.}
\end{figure}

\begin{figure}
\vspace{17truecm}
\caption{\small
Density contours in the orbital plane for the BH+NS merger
Models A5 (left column) and B5 (right column). The density (in g/cm$^3$)
is shown logarithmically, the contours being spaced with intervals of 
0.5~dex. The arrows indicate the velocity field. The time is given in 
the upper right corner of the plots.}
\end{figure}

\section{Results}

In this proceedings contribution the results of our models can 
only be outlined.
The results pertaining to the accretion of a high-density torus by a
black hole can be found in Ruffert \& Janka~(1999). Latest models
for neutron star merger simulations (the ones discussed here) will be 
published fully in Ruffert \& Janka~(2001, in preparation),
older calculations (obtained with a less complete 
version of our code) were published by Ruffert et al.\ 1996, 1997),
while the BH+NS simulations are summarized in Janka et al.\ (1999) and
were reported in detail by Eberl (1998).

\subsection{Dynamical Evolution and Mass Ejection}

Figure~3 shows the temporal evolution of the density
distribution in the orbital plane for both Models~A64 and~B64.
Initially, the orbits of the two neutron stars decay due to
gravitational radiation, and the stars approach each other slowly.
Once the distance decreases below the instability limit, the final plunge
occurs (upper panels).
Note that Model~B64, which has more total angular momentum, develops
prominent spiral arms in which matter is flung out to large
distances (right middle panel) and a considerable amount of matter 
carries enough mechanical energy to escape the system
($\Delta M_{\mathrm{ej}}\sim 2\times 10^{-2}\,$M$_{\odot}$, see Table~1).
Model~A64, on the contrary, remains more compact during this very early
post-merging phase (left middle panel). The amount of ejected mass for 
the A-model is therefore about one order of magnitude lower.
After 10~ms a very dense, nearly spherical central object 
has formed, surrounded by a less dense, extended, thick equatorial torus. 
When a steady state has been reached, the densities in this cloud are
rather similar in both simulations (lower panels).

With a Newtonian code we cannot determine whether and if, when, the 
merger remnant collapses to a black hole. In order to investigate what 
happens in such a case, we continued Run~B64, assuming that such a
catastrophe happens to the central, dense body on a dynamical timescale
of a few milliseconds after the merger. In the simulation listed in Table~1
(Model TN10) this gravitational instability was assumed to occur at the 
end of the computed merger evolution at $t = 10\,$ms. A Newtonian potential,
\begin{equation}
\Phi_{\rm N}\,=\,-{GM_{\rm BH}\over r} \ ,
\end{equation}
was taken for the black hole. In order to include the effects associated
with the existence of an innermost stable circular orbit at a radius of
$3R_{\rm s} = 6GM_{\rm BH}/c^2$, we compared these results with a second
(set of) simulation(s), in which the gravitational potential was described
by the pseudo-Newtonian expression,
\begin{equation}
\Phi_{\rm PW}\,\equiv\,-{GM_{\rm BH}\over r-R_{\rm s}}
\end{equation}
(Paczy\'nski \& Wiita 1980). 

The black hole at the center of the computational grid was represented by 
a vacuum sphere with a radius of $2R_{\rm s}$ instead of $1R_{\rm s}$ as 
in the BH+NS merger simulations. The gain of mass, momentum,
angular momentum and energy by the black hole due to the accretion of gas 
were also monitored during the simulations. In the grid zones
that were located inside the vacuum sphere (these zones were not
removed from the hydrodynamic grid), the mass density was
continuously reset to a negligibly small but finite value of
$10^8\,{\rm g\,cm}^{-3}$, and a correspondingly very small value of
the pressure was present.

After $\sim 5\,$ms of simulated accretion Model~TN10 has reached a 
quasi-stationary state
where the black hole has a mass of 2.9$\,$M$_{\odot}$ and the 
accretion torus contains $M_{\mathrm{d}}\sim 0.26\,$M$_{\odot}$ of gas. 
The density structure at this time is shown in  Fig.~4.
The left plot gives the density in the equatorial plane,
the right one shows it in the $x$-$z$- and $y$-$z$-planes perpendicular
to the equatorial plane. The contour for density 
$\rho = 10^{10}\,{\rm g\,cm}^{-3}$ extends out to a radius of 120~km, and
the average density of the torus is about $3\times 10^{11}\,{\rm g\,cm}^{-3}$.
At the end of the simulation the torus has
become nearly axially symmetric with only minor deviations. The accretion 
rate by the black hole ist $\dot M\sim 5$~M$_{\odot}\,$s$^{-1}$,
which allows one to estimate the duration of this accretion phase to 
$t\sim 53\,$ms. The evolution is driven by angular momentum
transport/loss due to numerical viscosity, corresponding to an effective
$\alpha$-viscosity of $\alpha_{\mathrm{eff}}\sim 4\times 10^{-3}$
(Table~2 and Ruffert \& Janka 1999).

{\small
\begin{table}[t]
\tabcolsep 2pt
\begin{center}
\caption{Accretion phase and neutrino annihilation}
{\small
\begin{tabular}{lcrrrcccccrrrr}
\hline\hline\\[-3mm]
Model$\,$ & 
$M_{\rm d}$  & $\,\dot M_{\rm d}\,$ & $\,\,t_{\rm acc}\,\,$ & $\,\,\alpha_{\rm eff}$ &
$\ \ \ \ a_{\rm i}\ \ $ & $\ \ a_{\rm f}\ \ $ & $\ \,a_{\rm BH}^\infty$ &
$\ave{L_{\nu}}$ & $\dot E_{\nu\bar\nu}$ & $q_{\nu}$ & $q_{\nu\bar\nu}$ & 
$E_{\nu}$ & $E_{\nu\bar\nu}$\\
   &   
   $M_{\odot}$ & $M_{\odot}/{\rm s}$ & ms & $10^{-3}$ & 
&   &   & $100\,{{\rm foe}\over{\rm s}}^a$ &
foe/s & \% & \% &  foe & foe\\[1mm]
\hline\\[-3mm]
   S64   & ... & ... & ... & ... & 0.98 & 0.75 & ... & 1.5 &  1  & ... & 1    & ... & ... \\
   D64   & ... & ... & ... & ... & 0.87 & 0.69 & ... & 2   &  2  & ... & 1    & ... & ... \\
   C64   & ... & ... & ... & ... & 0.64 & 0.49 & ... & 4   &  9  & ... & 2    & ... & ... \\
   A64   & ... & ... & ... & ... & 0.76 & 0.55 & ... & 5   &  9  & ... & 2    & ... & ... \\
   B64   & ... & ... & ... & ... & 0.88 & 0.63 & ... & 3   &  7  & ... & 2    & ... & ... \\
\hline
   TN10  &0.26 & 5   & 53  & 4 & ... & 0.42 & 0.59 & 1.2 & 0.5 & 1.3 & 0.4 & 7  & 0.03 \\
\hline
   C2.5  & 0.26 & 6    & 43  & 4    & 0.65 & 0.47 & 0.60 &  7  & 20 & 6     & 3 & 30   &   0.9 \\
   A2.5  & 0.33 &$<14$ &$>24$& $<8$ & 0.67 & 0.39 & 0.56 &  7  & 20 & $>3$  & 3 & $>17$&$>0.5$ \\
   B2.5  & 0.45 &$<35$ &$>13$& $<14$& 0.69 & 0.38 & 0.61 &  7  & 20 & $>1$  & 3 & $>9$ &$>0.3$ \\
   C5    & 0.38 & 5    & 76  & 5    & 0.44 & 0.27 & 0.42 &  4  & 8  & 4     & 2 & 30   &   0.6 \\
   A5    & 0.49 & 6    & 82  & 4    & 0.45 & 0.17 & 0.37 &  4  & 8  & 4     & 2 & 33   &   0.7 \\
   B5    & 0.45 & 6    & 75  & 5    & 0.46 & 0.19 & 0.38 &  4  & 8  & 4     & 2 & 30   &   0.6 \\
   C10   & 0.67 &$<10$ &$>67$& $<11$& 0.24 & 0.07 & 0.25 &  2  & 2  & $>1$  & 1 & $>13$&$>0.1$ \\
   A10   & 0.56 &$<60$ &$>9$ & $<82$&0.25  & 0.07 & 0.22 &  2  & 2  & $>0.2$& 1 & $>2$ &$>0.02$\\
   B10   & 0.47 & 3    & 160 & 5    & 0.25 & 0.11 & 0.23 &  2  & 2  & 4     & 1 & 32   &   0.3 \\[1mm]
\hline
\end{tabular}
}
\end{center}
$^a$ 1 foe = $10^{51}\,$erg ({\bf f}ifty {\bf o}ne {\bf e}rg).
\end{table}
}
 
Near the poles of the black hole and along the system axis, the density
has decreased to values below $5\times 10^8\,{\rm g\,cm}^{-3}$. 
This is only one order of magnitude above the lower density limit which
was set to $5\times 10^7\,{\rm g\,cm}^{-3}$ in the surroundings of the 
torus for numerical reasons, but nevertheless it is about 3 orders 
of magnitude below the average density inside the torus.
In this sense we see the cleaning of the region along the rotational axis 
of the black hole-torus system into of an ``evacuated'', cylindrical funnel
with opening half-angle of roughly 50 degrees.
Material which was swept into the polar regions during and immediately 
after the merging of the neutron stars falls into the newly formed black hole
very quickly within a free-fall timescale, because it is not supported by
centrifugal forces. This offers favorable conditions for the formation of
a baryon-poor, jetted outflow, powered by neutrino-antineutrino annihilation. 

The evolution of the density distributions of the BH+NS merger Models~A5 and B5
is shown in Fig.~5. During its first approach, Roche lobe overflow occurs
(upper panels) and the neutron star transfers
matter to the black hole at huge rates of several 100 up to 
$\sim 1000\,$M$_{\odot}\,$s$^{-1}$.
Within 2--3$\,$ms it loses 50--75\% of its initial mass. In case of 
the 2.5$\,$M$_{\odot}$ black hole the evolution is catastrophic and the
neutron star is immediately disrupted (for details, see Janka et al.\ 1999). 
A mass of 0.2--0.3$\,M_{\odot}$ remains in a thick disk around the
black hole ($M_{\rm d}$ in Table~2). In contrast, the neutron star
survives and is kicked on an elliptical orbit 
for $M_{\rm BH} = 5\,$M$_{\odot}$ and 10$\,$M$_{\odot}$,
In this case the orbital distance increases again
and a significantly less massive neutron star begins a second
approach. Near the periastron passage, 
the black hole again swallows gas at rates of more than
$100\,M_{\odot}$/s. Even a third episode of mass transfer
is possible (middle panels). Finally, at a distance $d_{\rm ns}$ and 
time $t_{\rm ns}$ 
the neutron star with a mass of $M_{\rm ns}^{\rm min}$ is destroyed
(Table~1) and most of its mass ends up in an accretion torus
(Table~2, lower panels of Fig.~5). 

The mass tranfer is not stable Roche lobe overflow, but occurs in a catastrophic, 
episodic way. After at most a few approaches, the neutron star is finally 
torn apart by gravitational forces. 
Note that in case of a physical equation of state, different from the 
ideal gas case, the neutron star mass is bounded by a lower limit. Below 
this limit the effective adiabatic index becomes smaller than 4/3 and a 
stable configuation
is not possible. In case of cold neutron stars, the lower mass limit
is typically around 0.1--0.2$\,$M$_{\odot}$. If the neutron stars are hot
(dimensionless entropies between 1 and 2 per nucleon, corresponding to
temperatures up to a few 10 MeV) and/or lepton-rich, 
this lower mass limit can be as large as
0.8--1$\,$M$_{\odot}$ (see, e.g., Gondek, Haensel, \& Zdunik 2000). Although
in our simulations the temperatures do not become that high because there is
only little heating, the neutron
star masses at the moment of destruction are near this value. The reason for 
this is the tidal stretching of the stars, which reduces their central densities
below the critical density where the adiabatic index drops to values smaller 
than 4/3. This destabilizes the neutron stars and leads to their final
disruption. Therefore a realistic equation of state does not allow a large
number of mass transfer cycles.

During the merging of BH+NS systems a gas mass 
$\Delta M_{\rm ej}$ between $\sim 10^{-4}\,M_{\odot}$
in case of counter-rotation and $M_{\rm BH} = 2.5\,M_{\odot}$, and
$\sim 0.1\,M_{\odot}$ for corotation and
$M_{\rm BH} = 10\,M_{\odot}$ is dynamically ejected (Table~1). 
In the latter case the associated angular momentum loss is about 7\%,
in all other cases 
it is less than 5\% of the total initial angular momentum of the system.
Another fraction of up to 24\% of the initial angular momentum is carried
away by gravitational waves. In Table~2  the rotation parameter $a = Jc/(GM^2)$ is
given for the initial state of the binary system ($a_{\rm i}$) and at
the end of the simulation ($a_{\rm f}$) for the remnant of NS+NS mergers or
for the black hole in BH+NS systems, respectively, provided the black
hole did not have any initial spin. When the whole disk mass 
$M_{\rm d}$ has been swallowed by the black hole, which is spun up,
a final value $a_{\rm BH}^\infty$ can be estimated by assuming accretion of
a corotating, thin disk with maximum radiation efficiency (Table~2).

At the end of the simulations, several of the BH+NS models have
reached a steady state, characterized by only a ``slow'' growth of the
black hole mass with a nearly constant accretion rate. Corresponding
rates $\dot M_{\rm d}$ are given in Table~2 and are several
$M_\odot$ per second. With typical torus masses $M_{\mathrm{d}}$ of several 
tenths of a solar mass, we estimate torus life
times $t_{\rm acc} = M_{\rm d}/\dot M_{\rm d}$ of 50--150$\,$ms. 
Values with $>$ and $<$ signs indicate
cases where the evolution and emission are still strongly
time-dependent at the end of the computation. In these cases the
accretion torus around the black hole has also not yet developed axial
symmetry. In all other cases the further disk evolution is mainly driven by the 
angular momentum transport mediated by viscous shear forces, which determine
the accretion rate. In the simulations the viscosity is associated with the
finite chosen resolution of the hydro code (which solves the Euler equations),
instead of a physical mechanism which creates viscosity. The corresponding
$\alpha$ parameter can be estimated to be $\alpha_{\rm eff}\sim 5\times 10^{-3}$
(Table~2), not an unrealistic value for accretion disks.

\begin{figure}
\vspace{17truecm}
\caption{\small
Ray-tracing images of six stages during the merging of a neutron star with
a $5\,$M$_{\odot}$ black hole (Model A5). 
The corresponding times from top left to bottom right are 3.0~ms, 5.2~ms, 
6.3~ms, 10.7~ms, 16.5~ms and 20.0~ms after the start of the simulation.
The images visualize the neutrino emission.
Bright regions are dense and emit neutrinos from optically thick conditions.}
\end{figure}

\begin{figure}
\plottwo{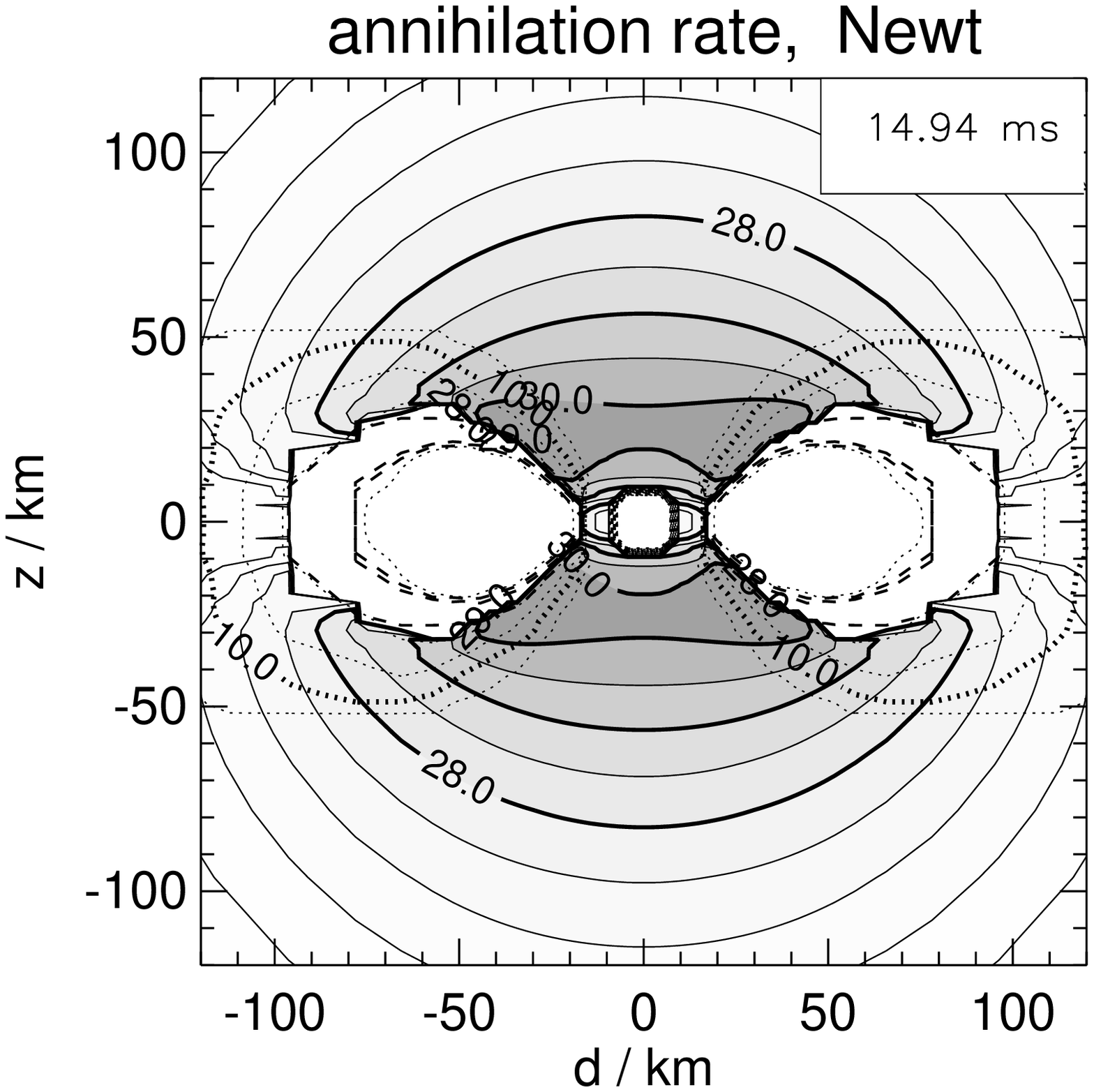}{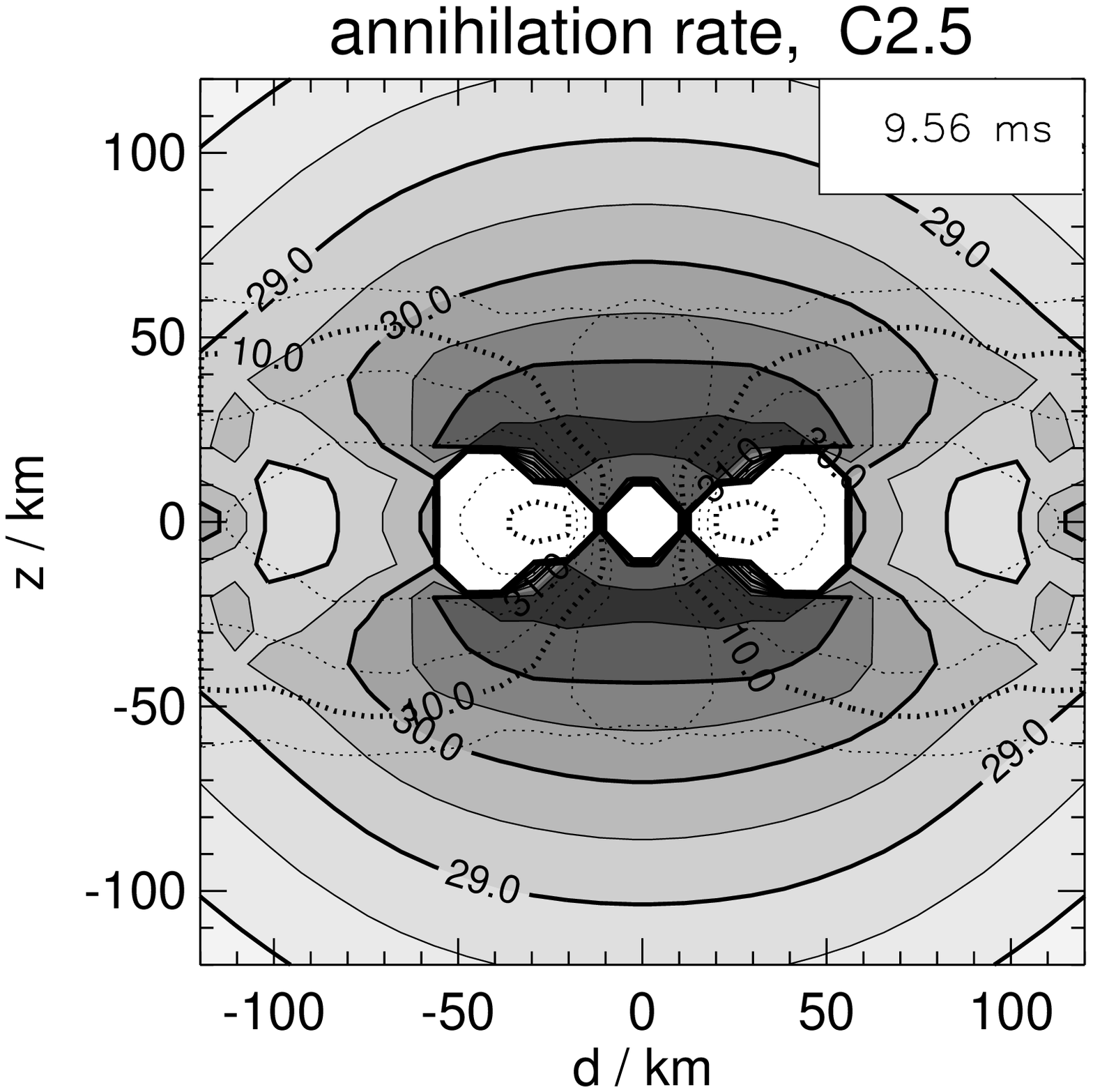}
\caption{\small
Energy deposition in the vicinity of a black hole by the annihilation
of neutrino-antineutrino pairs emitted from a hot accretion torus. The
left plot shows the situation after the NS+NS merging in Model~TN10, 
the right figure is the result of the BH+NS merger Model~C2.5. 
The white circles at the center represent the black hole, the solid
contours give the azimuthally averaged
energy deposition rate per unit volume (erg per cm$^3$ per s) and the
dashed contours show the density distribution of the tori.} 
\end{figure}

{\small
\begin{table}[t]
\tabcolsep 2.5pt
\begin{center}
\caption{Gravitational waves and neutrinos}
{\small
\begin{tabular}{lccccccccccc}
\hline\hline\\[-3mm]
Model & $L_{\rm GW}^{\rm max}$  & $rh^{\rm max}$ & $E_{\rm GW}$ &
$L_{\nu_e}^{\rm max(av)}\,$ & $\,L_{\bar\nu_e}^{\rm max(av)}\,$ 
& $\,L_{\Sigma\nu_x}^{\rm max(av)}\,\,$ & $E_{\nu}$ & 
$kT^{\rm max}$ & $\ave{\epsilon_{\nu_e}}$ 
& $\ave{\epsilon_{\bar\nu_e}}$ & $\ave{\epsilon_{\nu_x}}$\\[1mm]
      & $10^4\,{{\rm foe}\over{\rm s}}^a$ & $10^4{\rm cm}$ & foe &
$100\,{{\rm foe}\over{\rm s}}$ & $100\,{{\rm foe}\over{\rm s}}$ 
& $100\,{{\rm foe}\over{\rm s}}$ &
foe & MeV & MeV & MeV & MeV\\[1mm]
\hline\\[-3mm]
   S64   & 0.7 & 5.5 & 14 & 0.3(0.2) & 0.9(0.5) & 0.3(0.2) & 0.8 & 35 & 12 & 18 & 26\\
   D64   & 0.4 & 5.5 & 13 & 0.5(0.3) & 1.3(0.8) & 0.7(0.4) & 1.1 & 35 & 13 & 19 & 27\\
   C64   & 1.2 & 6.0 & 23 & 1.1(0.5) & 2.6(1.3) & 0.7(0.3) & 1.9 & 69 & 13 & 19 & 27\\
   A64   & 2.1 & 8.6 & 52 & 0.9(0.5) & 2.6(1.3) & 1.4(0.6) & 2.3 & 39 & 12 & 18 & 26\\
   B64   & 2.1 & 8.9 & 37 & 0.6(0.4) & 1.8(1.1) & 0.9(0.4) & 1.8 & 39 & 13 & 19 & 27\\
\hline
   TN10  & ... & ... & ... & 0.5(0.4) & 1.3(0.9) & 0.6(0.2) & 0.8 & 15 &  9 & 13 & 21\\
\hline
   C2.5  & 2.3 & 9.9  & 32  & 1.5(0.5) & 7.3(2.5) & 5.2(1.9) & 4.5 & 74 & 16 & 22 & 31\\
   A2.5  & 2.0 & 9.9  & 50  & 1.8(0.5) & 6.4(2.2) & 3.1(1.3) & 3.6 & 65 & 15 & 22 & 31\\
   B2.5  & 2.1 & 9.6  & 61  & 0.9(0.3) & 6.5(1.7) & 3.6(0.9) & 2.5 & 61 & 14 & 21 & 29\\
   C5    & 3.9 & 13.0 & 50  & 0.7(0.4) & 3.8(1.6) & 2.5(1.1) & 4.5 & 46 & 15 & 20 & 29\\ 
   A5    & 3.2 & 14.8 & 102 & 0.7(0.2) & 4.4(1.5) & 2.8(0.8) & 4.5 & 51 & 16 & 24 & 31\\
   B5    & 3.4 & 14.5 & 95  & 0.6(0.2) & 3.7(1.1) & 2.5(0.6) & 2.9 & 44 & 14 & 21 & 28\\
   C10   & 7.1 & 21.9 & 123 & 0.4(0.1) & 2.5(0.4) & 1.2(0.1) & 0.6 & 51 & 14 & 19 & 24\\
   A10   & 6.9 & 26.2 & 168 & 0.2(0.1) & 2.5(0.5) & 1.2(0.2) & 0.7 & 50 & 14 & 20 & 26\\
   B10   & 7.3 & 26.2 & 163 & 0.4(0.1) & 2.5(0.8) & 1.4(0.2) & 1.1 & 52 & 13 & 18 & 24\\[1mm]
\hline
\end{tabular}
}
\end{center}
$^a$ 1 foe = $10^{51}\,$erg ({\bf f}ifty {\bf o}ne {\bf e}rg).
\end{table}
}

\subsection{Neutrino Emission and Gamma-Ray Bursts}

Compressional heating, shear due to numerical viscosity, and dissipation 
in shocks heat the gas in the merging neutron stars and in the accretion
flow to the black hole. After a few milliseconds the maximum temperatures
in the NS+NS merger remnant have climbed to several 10$\,$MeV (Table~3).
This is also the typical timescale until the thermodynamic conditions in the 
accretion torus around the black hole reach a steady state with average
temperatures around 10$\,$MeV and densities between $10^{10}\,$g/cm$^3$ and
$10^{12}\,$g/cm$^3$. At such conditions electron and positron pairs are
abundant, and electron neutrinos and antineutrinos are copiously produced
by $e^\pm$ capture reactions on free nucleons
\begin{equation}
p + e^- \longrightarrow n + \nu_e\quad{\mathrm{and}} \quad 
n + e^+ \to p + \bar\nu_e\ .
\end{equation}
Muon and tau neutrino and antineutrino pairs are also emitted by $e^+e^-$ 
annihilation but 
contribute to the energy loss at a minor level. The accretion flow is not
transparent to neutrinos and effects due to their finite diffusion time have
to be taken into account when the neutrino luminosity is calculated.  
In Fig.~6 the thermal evolution of the BH+NS merger Model~A5 can be seen
from ray-tracing images which visualize the neutrino emission. 

Table~3 provides information about the maximum and time-averaged 
luminosities ($L_{\nu_i}^{\rm max}$ and $L_{\nu_i}^{\rm av}$,
respectively) for $\nu_e$ and $\bar\nu_e$ and for the sum of all
heavy-lepton neutrinos (which are denoted by 
$\nu_x \equiv \nu_\mu,\,\bar\nu_\mu,\,\nu_\tau,\,\bar\nu_\tau$).
The total neutrino emission fluctuates strongly in response to
dynamical processes, in particular rises dramatically during mass 
transfer episodes in BH+NS mergers. In case of NS+NS mergers there is a 
more monotonic increase with time. The average energies of emitted 
neutrinos, $\ave{\epsilon_{\nu}}$, are typically about 15$\,$MeV for
electron neutrinos, about 20$\,$MeV for electron antineutrinos and 
about 30$\,$MeV for muon and tau neutrinos and antineutrinos. 
The total energy $E_\nu$ radiated in neutrinos within the computed
10--20$\,$ms of the evolution is a few $10^{51}\,$erg.
In case of BH+NS mergers, in particular for smaller black holes,  
the neutrino luminosities and mean energies are significantly higher 
than for NS+NS mergers. 

The black hole continues to swallow matter from the accretion 
torus for a period between several ten milliseconds and possibly 
a fair fraction of a second (Table~2). During this time the hot gas
will continue to radiate neutrinos. We cannot follow this long-time 
evolution by hydrodynamical models, but attempt to derive rough 
estimates by extrapolating the conditions present at the end of our 
simulations. With average total neutrino luminosities $\ave{L_\nu}$
of several $10^{53}$~erg$\,$s$^{-1}$ the accretion tori lose a total
energy of a few $10^{52}\,$erg in neutrinos ($E_{\nu}$ in Table~2),
which equals up to 4--6\% of the rest-mass energy of the 
accreted matter [$q_{\nu}\equiv E_{\nu}/(M_{\rm d}c^2)$]. 
Because the BH-torus configuration is very compact and neutrinos 
are emitted at very high luminosities, the annihilation of 
neutrino-antineutrino pairs, $\nu\bar\nu \to e^+e^-$, 
in the vicinity of the black hole is 
rather efficient, depositing an energy $E_{\nu\bar\nu}$ of nearly
$10^{51}\,$erg (Table~2) in a pair-plasma cloud 
above the poles of the black hole
(Fig.~7). Peak rates of this energy deposition exceed $10^{52}\,$erg/s.
The efficiency of energy conversion from neutrinos to $e^\pm$ pairs,
$q_{\nu\bar\nu} \equiv E_{\nu\bar\nu}/E_{\nu}$, is of the order of
1--3 per cent (Table~2).

\begin{figure}
\vspace{17truecm}
\caption{\small
Gravitational waveforms for NS+NS merger Models~A64, B64 and C64 and 
all BH+NS merger models listed in Table~1. Time is measured in ms from
the start of the simulation, the observer is located perpendicular to 
the orbital plane. The thin solid lines correspond to the chirp
signal of two point masses.} 
\end{figure}

\subsection{Gravitational-Wave Emission}

While the neutrino emission rises gradually after the final plunge of
binary neutron stars, the gravitational-wave emission peaks at the 
moment of the collision of the stars. In case of BH+NS mergers the 
maximum gravitational-wave luminosity coincides with the phase when the
mass transfer rate to the black hole is largest, which occurs
during the first approach of the neutron star (between 2 and 5$\,$ms
after the start of the simulations). Afterwards the neutron
star is either disrupted to an essentially axi-symmetric accretion torus 
(in case of a 2.5$\,$M$_{\odot}$ black hole) or has been stripped
to a significantly less massive body. Both reduces the production of 
gravitational waves. The maximum gravitational-wave luminosity 
$L_{\mathrm{GW}}^{\mathrm{max}}$ and the wave amplitude $rh$ (for distance
$r$ from the source) increase with the black hole mass and with the 
masses of the merging neutron stars (Table~3). 
$L_{\mathrm{GW}}^{\mathrm{max}}$ can reach values of nearly 
$10^{56}$~erg$\,$s$^{-1}$ in case of 10$\,$M$_{\odot}$ black holes, and the
total energy radiated in gravitational waves, $E_{\rm GW}$, can be as high
as $0.1\,$M$_{\odot}c^2$. Fourier spectra of the emitted waves peak between
800 and 1000~Hz, with the dominant frequency of the emission decreasing 
with the black hole mass.

Figure~8 shows typical waveforms for NS+NS and BH+NS mergers. The numerical
simulations reproduce the chirp signal of inspiralling point masses very
well before the dynamical interaction of the compact objects sets in. 
The final plunge of the two neutron stars produces maxima of the wave 
amplitude, the ongoing wave emission after the merging reflects the 
oscillations and the ringing of the massive remnant. The detailed structure
of the signal during this phase depends on the assumed neutron star spins.
If the remnant collapses to a black hole, this emission would be cut off.

For BH+NS mergers significant emission continues for a much longer
time than predicted by the point-mass approximation, especially in case of
5$\,$M$_{\odot}$ and 10$\,$M$_{\odot}$ black holes where the neutron
star survives the first approach, although it loses a major fraction of its
mass to the black hole. The emission decreases to a negligible level once the
neutron star is torn apart and its gas forms an accretion torus around the
black hole. Maximum wave amplitudes $rh$ are between $\sim 6\times 10^4\,$cm
and $\sim 30\times 10^4\,$cm (Table~3).

\section{Conclusions}

Our three-dimensional hydrodynamical simulations of NS+NS and BH+NS merging,
although basically Newtonian, allow for some interesting conclusions on 
potentially observable implications. We find that during the dynamical
phase of the merging gas with a mass between $\sim 10^{-4}\,$M$_{\odot}$ and
several $10^{-2}\,$M$_{\odot}$ may be ejected from the system and may 
become unbound. The exact amount depends sensitively on the neutron star spin(s)
and on the mass of the black hole. The ejected gas may contribute to the chemical
composition of the interstellar gas at a significant level. The nuclear composition
of this gas has to be determined by detailed calculations of radioactive
decays and possible r-processing. First interesting results have been 
obtained (Freiburghaus et al.\ 1999), but more work and more models are necessary.
The ejection of radioactive material might also
produce transient visible radiation, which could be observed in distant galaxies
(Li \& Paczy\'nski 1998).

A detection of gravitational waves from the final, dynamical phase of the 
coalescence of NS+NS and BH+NS systems would yield 
rich information about the masses of the compact objects, their radii and
the equation of state of neutron star matter. The upcoming km-scale interferometers 
have a chance to make such a ground-braking measurement, although estimates of
the probability are very uncertain due to the observationally undetermined rates of  
mergers and many unknows in theoretical rate predictions. In addition, the 
sensitivity of experiments like LIGO is maximal around 100~Hz, and is worse by
about a factor of 10 at 1000~Hz. The signal from the final plunge might be 
observable with LIGO only to a distance of $\sim 1\,$Mpc, and with the advanced
LIGO instrument only out to $\sim 10\,$Mpc. 

Whereas gravitational waves trace the geometry and mass distribution 
of the coalescing compact objects 
and the dynamics of the merging process, neutrinos can provide us with information
about the thermodynamics of the gas. The emission of MeV neutrinos from NS+NS or
BH+NS merger events could be comparable in energy to the neutrino signal from a 
supernova (in particular, if the NS+NS merger remnant does not collapse to a black
hole quickly). Nevertheless, a direct measurement of these neutrinos
is extremely unlikely, because the most optimistic estimates expect only one event
in our Galaxy within $10^4$ years, more than 100 times less frequent than core-collapse
supernovae. Measuring the MeV neutrinos from extragalactic mergers will not be 
possible in the near future. If the catastropic destruction of NS+NS or BH+NS
systems is associated with the production of a gamma-ray burst, however, 
high-energy in the 100~TeV range (Waxman \& Bahcall 1997) 
and even ultra high-energy ($\sim 10^{18}$~eV) neutrinos (Waxman \& Bahcall 2000) 
might be produced, which could lead to events in future km$^3$ detectors like 
AMANDA in the Antarctica.

We explored the possibility of powering a GRB by neutrino-antineutrino annihilation
in the vicinity of the black hole, which accretes matter after the merging of the
compact objects. Typical torus masses were found to be several tenths of a solar
mass, with accretion rates of a few solar masses per second. The lifetime of the
accretion disk might therefore be fractions of a second. During this time neutrinos
could deposit an energy up to $\sim 10^{51}$~erg in a $e^\pm$ plasma cloud in
the baryon-poor funnel above the poles of the black hole. Currently it is not
clear how much these estimates may be affected by different properties of the
accretion torus in a general relativistic simulation and by
the rotation of a Kerr black hole. A possibly
ultrarelativitically expanding jet would not be strongly collimated, but opening
half-angles of several 10 degrees could increase the equivalent isotropic energy 
inferred by an observer in jet direction to values of several $10^{52}\,$erg.
The observable GRBs would probably have a hard spectrum and would be luminous above
average (because of high $\nu\bar\nu$ energy deposition rates and not very large 
cosmological redshift), but would 
be of the short-duration type and less energetic than the long bursts. If the 
jet has to clean the axial funnel from baryons before maximal Lorentz factors
are reached, a softer precursor to the actual GRB may be possible. 
Since the gas densities in the merger environment are low, the occurrence
of afterglows seems unlikely, but matter, ejected during the merging or from 
the accretion torus by a neutrino-driven wind or by magnetic field effects might
lead to associated emission at other wavelengths.

Very important questions cannot be satisfactorily answered by our current
models but require further, preferably general relativistic simulations. 
Does the remnant of a 
NS+NS merger collapse to a black hole and if so, when does it collapse?
Is the ejection of mass found in Newtonian models also present in case of 
the stronger relativistic gravity? How much mass ends up in an accretion 
torus around the black hole in this case? How does general relativity change the 
properties, neutrino emission and neutrino-antineutrino annihilation 
of the accretion disk? What are the lifetime of the accretion torus and 
how much energy is emitted by neutrinos or released by magnetic fields?
Can a baryon-poor outflow form along the rotation axis and if so, how
strong is the collimation of such a pair-plasma jet? What are the effects of 
a rapid rotation of the black hole? Do Kerr effects raise the efficiency
of energy conversion from accretion to neutrinos and further to $e^\pm$
plasma? These questions have only been touched by current investigations
and much more work is desirable. We have started to take a next, small 
step by using a pseudo-Newtonian (Paczy\'nski \& Wiita 1980) potential for the 
black hole in BH+NS systems, which allows us to roughly include the effects
due to the existence of an innermost stable circular orbit. Using the 
description by Artemova, Bjoernsson, \& Novikov (1996) we also consider the 
influence of the Kerr character of the black hole on the radius of the innermost
stable orbit.

\section*{Acknowledgements}
HTJ and MR would like to thank the organizers for an exciting and
perfectly managed conference.
Work on this project was supported by the U.K.~PPARC as
Advanced Fellow for MR and by the Deutsche Forschungsgemeinschaft on grant 
``Sonderforschungsbereich 375 f\"ur Astro-Teilchenphysik'' for HTJ.
We thank T.~Eberl for running the BH+NS merger simulations as
work for his Diploma Thesis.
The calculations were performed at the Rechenzentrum Garching
of the Max-Planck Gesellschaft.


\begin{references}          

\reference
Artemova, I.V., Bjoernsson, G., \& Novikov, I.D. 1996, \apj, 461, 565

\reference
Balberg, S., \& Shapiro, S.L. 2000, astro-ph/0004317 

\reference
Baumgarte, T.W., \& Shapiro, S.L. 1998, \apj, 504, 431

\reference
Baumgarte, T.W., Shapiro, S.L., \& Shibata, M. 2000, \apj, 528, L29

\reference
Berger, M.J. 1987, J. Numer. Anal., 24, 967

\reference
Berger, M.J., \& Colella, P. 1989, J. Comput. Phys., 82, 64

\reference
Berger, M.J., \& Oliger, J. 1984, J. Comput. Phys., 53, 484

\reference
Blanchet, L., Damour, T., \&  Sch\"afer, G. 1990, \mnras, 242, 289

\reference
Blandford, R.D., \& Znajek, R.L. 1977, \mnras, 179, 433

\reference
Blinnikov, S.I., Novikov, I.D., Perevodchikova, T.V., \& Polnarev, A.G.
1984, Sov. Astron. Lett., 10, 177

\reference
Bombaci, I., \& Datta, B. 2000, \apj, 530, L69

\reference
Bulik, T., Belczynski, K., \& Zbijewski, W. 1999, \mnras, 309, 629

\reference
Colella, P., \& Woodward, P.R. 1984, J. Comput. Phys., 54, 174

\reference
Duez, M.D., Baumgarte, T.W., \& Shapiro, S.L. 2000, gr-qc/0009064

\reference
Eberl, T. 1998, Diploma Thesis, Technical University Munich

\reference
Eichler, D., Livio, M., Piran, T., \& Schramm, D.N. 1989, Nature, 340, 126

\reference
Freiburghaus, C., Rosswog, S., \& Thielemann, F.-K. 1999, \apj, 525, L121

\reference
Fryer, C.L., Woosley, S.E., \& Hartmann, D. 1999, \apj, 526, 152

\reference
Gondek, D., Haensel, P., \& Zdunik, J.L. 2000, astro-ph/0012541

\reference
Heiselberg, H., \& Pandharipande, V. 2000, Ann. Rev. Nucl. Part. Sci., 50, 481

\reference
Hulse, R.A., \& Taylor, J.H. 1975, \apj, 195, L51

\reference
Janka, H.-T., \& Ruffert, M. 1996, \aap, 307, L33

\reference
Janka, H.-T., Eberl, T., Ruffert, M., \& Fryer, C.L. 1999, \apj, 527, L39

\reference
Kalogera, V., \& Lorimer, D.R. 2000, \apj, 530, 890

\reference
Klose, S. 2000, astro-ph/0001008

\reference
Klu\'zniak W., \& Ruderman, M. 1998, \apj, 505, L113

\reference
Kobayashi, S., \& Sari, R. 2000, astro-ph/0101006

\reference
Lamb, D.Q. 2000, astro-ph/0005028

\reference
Lattimer, J.M., \& Prakash, M. 2000a, Phys. Rept., 333--334, 121

\reference
Lattimer, J.M., \& Prakash, M. 2000b, astro-ph/0002232

\reference
Lattimer, J.M., \& Schramm, D.N. 1974, \apj, 192, L145

\reference
Lattimer, J.M., \& Schramm, D.N. 1976, \apj, 210, 549

\reference
Lattimer, J.M., \& Swesty, F.D. 1991,  Nucl. Phys., A535, 331

\reference
Lattimer, J.M., Mackie, F., Ravenhall, D.G., \& Schramm, D.N. 1977, 
\apj, 213, 225

\reference
Lee, W.H. 2001, astro-ph/0101132

\reference
Lee, W.H., \& Klu\'zniak, W. 1999a, \apj, 526, 178

\reference
Lee, W.H., \& Klu\'zniak, W. 1999b, \mnras, 308, 780 

\reference
Li, L.-X., \& Paczy\'nski, B. 1998, \apj, 507, L59

\reference
Li, L.-X., \& Paczy\'nski, B. 2000, \apj, 534, L197

\reference
Lipunova, G.V., \& Lipunov, V.M. 1998, \aap 329, 29

\reference
Mao, S., Narayan, R., \& Piran, T. 1994, \apj, 420, 171

\reference
Meyer, B.S. 1989, \apj 343, 254

\reference
Mitra, A. 2000, astro-ph/0010311

\reference
Narayan, R., Piran, T., \& Shemi, A. 1991, ApJ, 379, L17

\reference
Otsuki, K., Tagoshi, H., Kajino, T., \& Wanajo, S. 2000, \apj, 533, 424

\reference
Paczy\'nski, B. 1991, Acta Astronomica, 41, 257

\reference
Paczy\'nski, B., \& Wiita, P.J. 1980, \aap, 88, 23

\reference
Qian, Y.-Z., \& Woosley, S.E. 1996, \apj, 471, 331

\reference
Rosswog, S., Davies, M.B., Thielemann, F.-K., \& Piran, T. 2000,
\aap, 360, 171

\reference
Rosswog, S., Liebend\"orfer, M., Thielemann, F.-K., Davies, M.B., 
Benz, W., \& Piran, T. 1999, \aap 341, 499

\reference
Ruffert, M. 1992, \aap, 265, 82

\reference
Ruffert, M., \& Janka, H.-T. 1998, \aap, 338, 535

\reference
Ruffert, M., \& Janka, H.-T. 1999, \aap, 344, 573

\reference
Ruffert, M., Janka, H.-T., \& Sch\"afer, G. 1996, \aap, 311, 532

\reference
Ruffert, M., Janka, H.-T., Takahashi, K., \& Sch\"afer, G. 1997, \aap, 319, 122

\reference
Shibata, M., \& Ury\=u, K. 1999, astro-ph/9911058

\reference
Sumiyoshi, K., Yamada, S., Suzuki, H., \& Hillebrandt, W. 1998, \aap, 334, 159

\reference
Sumiyoshi, K., Suzuki, H., Otsuki, K., Terasawa, M., \& Yamada, S. 2000,
PASJ, 52, 601

\reference
Thorsett, S.E., \& Chakrabarty, D. 1999, \apj, 512, 288 

\reference
Ury\=u, K., Shibata, M., \& Eriguchi, Y. 2000, gr-qc/0007042

\reference
Usov, V.V. 1994, \mnras, 267, 1035

\reference
Waxman, E., \& Bahcall, J.N. 1997, Phys. Rev. Lett., 78, 2292

\reference
Waxman, E., \& Bahcall, J.N. 2000, \apj, 541, 707

\reference
Weber, F. 1999, Pulsars as Astrophysical Laboratories for Nuclear and 
Particle Physics (Bristol: IOP) 

\end{references}
\end{document}